\documentclass[%
twocolumn,
longbibliography,
 amsmath,amssymb,
 aps,
prl,
]{revtex4-1}

\usepackage{graphicx}% Include figure files
\usepackage{dcolumn}% Align table columns on decimal point
\usepackage{bm}% bold math
\usepackage{subfigure}
\begin{document}

\preprint{APS/123-QED}

\title{Quantum Hall Effect  in Hydrogenated Graphene }% Force line breaks with \\

\author{J. Guillemette$^{1,2}$,  S.S.  Sabri$^{2}$, B. Wu$^{1}$, K. Bennaceur$^{1}$, P.E. Gaskell$^{2}$, M. Savard$^{1}$, P.L. L\'evesque$^{3}$, F. Mahvash$^{2,4}$, A.  Guermoune$^{2,4}$, M.  Siaj$^{4}$, R. Martel$^{3}$, T. Szkopek$^{2\dag}$, and G.  Gervais$^{1\dag \star}$}

\affiliation{$^{1}$Department of Physics, McGill University, Montr\'{e}al, QC,  H3A 2T8, Canada}%Lines break automatically or can be forced with \\

\affiliation{$^{2}$ Department of Electrical and Computer Engineering, McGill University, Montr\'{e}al, QC, H3A 2A7, Canada}%Lines break automatically or can be forced 
\affiliation{$^{3}$ Department of Chemistry, Universit\'e de Montr\'eal, Montr\'{e}al, QC H3C 3J7, Canada}%Lines break automatically or can be forced 

\affiliation{$^{4}$ Department of Chemistry, Universit\'e du Qu\'ebec \`a Montr\'eal, Montr\'{e}al, QC, H3C 3P8, Canada}%Lines break automatically or can be forced \\

\affiliation{$^\dag$both authors contributed equally to this work}
\affiliation{$^\star$corresponding author: gervais@physics.mcgill.ca}
%\date{\today }

%\email{gervais@physics.mcgill.ca}
\date{\today}% It is always \today, today

\begin{abstract}

The quantum Hall effect is observed in a two-dimensional electron gas formed in millimeter-scale hydrogenated graphene, with a mobility less than 10 $\mathrm{cm^{2}/V\cdot s}$ and corresponding  Ioffe-Regel disorder parameter $(k_{F}\lambda)^{-1}\gg1$. In zero magnetic field and low temperatures, the hydrogenated graphene is insulating with a two-point resistance of order of $250 h/e^2$. Application of a strong magnetic field generates a negative colossal magnetoresistance, with the two-point resistance saturating within 0.5\% of $h/2e^{2}$ at 45T. Our observations are consistent with the opening of an impurity-induced gap in the density of states of graphene. The interplay between electron localization by defect scattering and magnetic confinement in two-dimensional atomic crystals is discussed.

\end{abstract}

\maketitle

%\tableofcontents

%\section{Introduction}

Two-dimensional atomic crystals have attracted much attention as surfaces with unique low-dimensional electron transport behaviour, most notably the zero-gap semiconductor graphene \cite{Novoselov04}. The relativistic Dirac dispersion of electrons in graphene leads to an anomalous four-fold degenerate quantum Hall (QH) sequence \cite{Novoselov05,Zhang05,Zhang06,Young12}. Common to these studies is low disorder, with a mean free path $\lambda$ large compared to electron Fermi wavelength $\lambda_F = 2\pi/k_F$, with the cleanest graphene samples exhibiting the fractional quantum Hall (FQH) effect \cite{Du09,Dean11}. We report here our discovery of a quantum Hall effect in graphene at the opposite extreme, whereby hydrogenation was used to induce a short mean-free path beyond the Ioffe-Regel limit $k_{F}\lambda\sim1$ for the onset of strongly insulating behaviour. Our work illustrates the importance of the interplay between electron localization by point defect scattering and magnetic confinement in two-dimensional atomic crystals.

\begin{figure}
\includegraphics[scale=1]{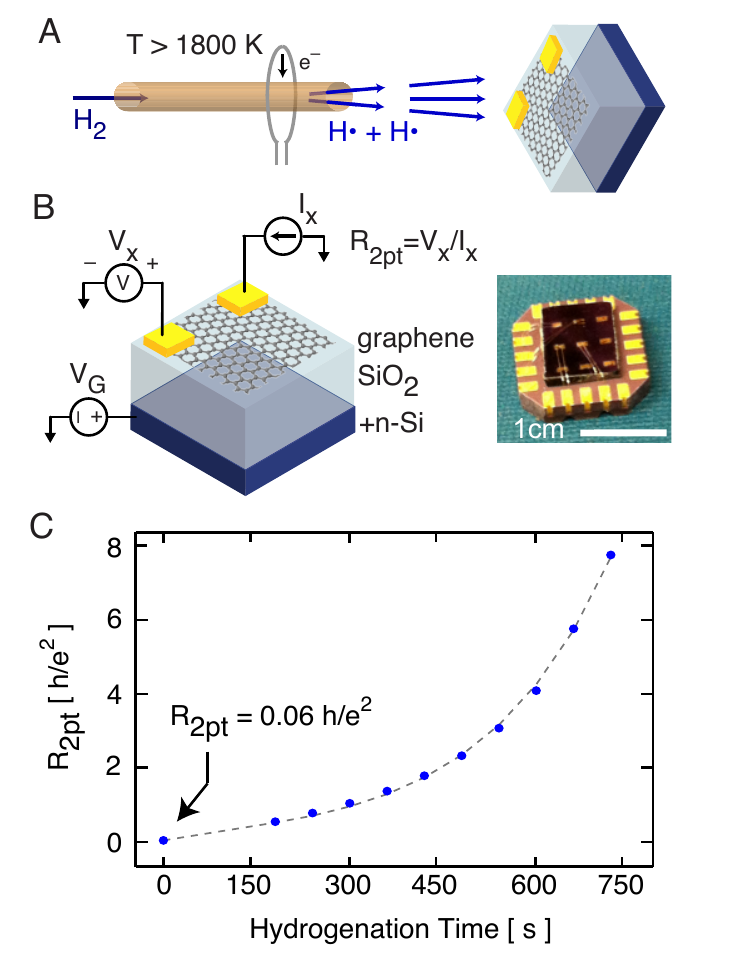}
\caption{Schematic of graphene hydrogenation. {\bf A)} An electrically contacted CVD grown graphene sample is exposed to reactive hydrogen atoms produced by thermal cracking of hydrogen at $T>$1800K in a UHV environment. {\bf B)} The two-point resistance $R_{2pt}$ of the graphene sheet is measured {\it in situ} with hydrogen dosing. A photograph of the device is shown. {\bf C)} The typical evolution of room temperature resistance versus hydrogen exposure time (circles), and an exponential fit to resistance versus hydrogen dose (dotted line). }
\label{fig:fig1}
\end{figure}

 We present experimental results on electronic transport in graphene where neutral point defects are introduced by hydrogenation. Previous electron transport studies of hydrogenated graphene were interpreted as evidence for graphane\cite{Elias09}, a two-dimensional polymer of carbon and hydrogen. More recently, angle resolved photo-emission spectroscopy (ARPES) \cite{Grueneis10} and scanning tunneling microscopy (STM) \cite{Grueneis12} measurements have demonstrated the opening of a gap in the density of states of hydrogenated graphene. The object of our work is to probe the magnetotransport properties of hydrogenated graphene. We focus on large area (millimeter-scale) graphene that was exposed to an atomic hydrogen beam resulting in insulating behaviour ($dR/dT<0$ at zero magnetic field) and a low-temperature resistance $R\sim250\times h/e^2$ far above the Ioffe-Regel limit for the onset of insulating behaviour. Application of a $45~\mathrm{T}$ magnetic field induces a transition to a well formed quantum Hall state with a two point resistance $R_{2pt}=h/2e^2$ accurate to within $0.5\%$. The insulator-quantum Hall transition coincides with the magnetic length ($\ell_{B}$)  approaching, and becoming less than the mean defect spacing inferred from Raman spectroscopy ($\lambda_{D}$).

\begin{figure}
\includegraphics[scale=0.4]{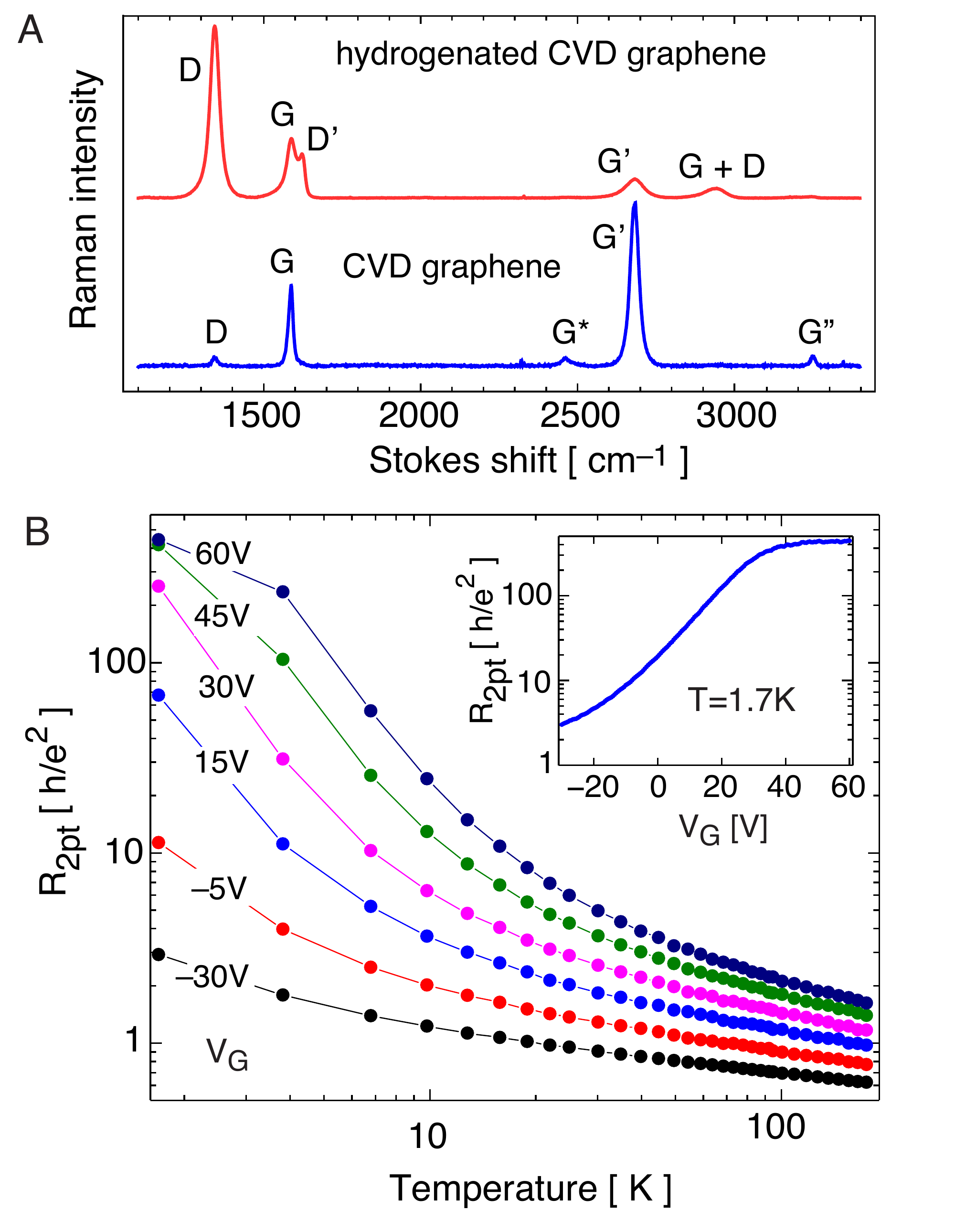}
\caption{Characterisation of the hydrogenated graphene devices. {\bf A)} Raman Stokes spectrum of CVD graphene (lower trace) and hydrogenated CVD graphene (higher trace). Peaks associated with $sp^2$ carbon are classified $G$, and peaks associated with disorder are classified $D$, respectively. The ratio $I_D/I_G$ of $D$ peak intensity to $G$ peak intensity is $2.2$ and $0.15$ for hydrogenated CVD graphene and pristine CVD graphene, respectively, with the corresponding mean defect spacing $\lambda_D$ estimated to be $4.6\pm0.5~\mathrm{nm}$ and $\sim35~\mathrm{nm}$ (see text). {\bf B)} The temperature dependence of the resistance is insulating ($dR_{2pt}/dT<0$) at all gate voltages. The field effect (inset) shows hole conduction and a zero field effect mobility regime at low hole density. }
\label{fig:fig2}
\end{figure}

Disordered graphene samples were prepared from pristine, large-area, monolayer graphene samples grown by chemical vapour deposition (CVD) on Cu foils \cite{Ruoff09}, with growth details reported elsewhere \cite{guermoune11}. Graphene monolayers were transferred to oxidized silicon wafers and electrically contacted for electron transport experiments.  Disorder was controllably introduced into the graphene by exposure to a beam of atomic hydrogen produced by thermally cracking molecular hydrogen with a white-hot tungsten capillary \cite{Bischler93}. The room temperature two-point resistance $R_{2pt}$ of graphene samples was monitored versus atomic hydrogen exposure {\it in situ} in an ultra-high vacuum chamber, Fig.\ref{fig:fig1}. The room temperature resistance of graphene devices was observed to increase with exposure to atomic hydrogen, as has been observed in graphene exposed to hydrogen plasma \cite{Elias09}. Atomic hydrogen exposure produces an exponential growth in resistance, and a corresponding exponential shortening of carrier mean-free path $\lambda$. The Ioffe-Regel disorder parameter $(k_{F}\lambda)^{-1} \simeq (M/2)\cdot(e^2/h)\cdot R_{2pt}$, where $M = 4$ accounts for spin and valley degeneracy, is consequently significantly increased beyond the limit for the onset of strongly insulating behaviour.

%The increase in resistance cannot be accounted for by doping via charged species because doping would have the effect of ultimately reducing the resistance, and doping would not produce a response exponential in dose.

The disorder introduced by hydrogenation was observed by Raman spectroscopy, Fig. \ref{fig:fig2}{\bf A}. The Raman Stokes peaks G and G' associated with pristine $sp^2$ carbon indicate clean, monolayer CVD graphene prior to hydrogenation (blue spectrum). The emergence of strong disorder-induced peaks D and D' upon hydrogenation (red spectrum) confirms the introduction of point defects that break the translational invariance of the crystal \cite{Jorio11}. The nature of the point defects introduced by hydrogenation is most likely hydrogen-carbon bonds \cite{Elias09}. By direct comparison of the ratio of D peak intensity to G peak intensity, $I_D/I_G$, with that observed in graphene disordered by controlled ion bombardment \cite{Lucchese10}, we estimate the mean hydrogen-hydrogen spacing to be in the range $\lambda_{D}\simeq 4.6\pm0.5~\mathrm{nm}$ for a 360s hydrogen exposure, corresponding to a $0.2-0.3\%$ hydrogen to carbon ratio.

%Richard & Pierre: Electron transport studies were performed on a graphene sample exposed to atomic hydrogen for 300s, corresponding to an estimated dose of <<< MAKE H DOSE ESTIMATE FROM FLOW RATE? >>> atomic hydrogen. 

Electronic transport measurements were performed on the hydrogenated graphene with large ohmic contacts at the opposing corners of a 2.5 mm square using a standard low-frequency technique. The hydrogenated graphene shows unambiguous insulating behaviour, $\partial R_{2pt}/\partial T < 0$, over the entire range of accessible temperatures (from 175K to $\sim$1.7K) and gate induced carrier density, Fig.\ref{fig:fig2}{\bf B}. Gate voltage modulation of the resistance shows hole transport with a field effect mobility $\mu = \partial (1/R_{2pt}) / \partial (C V_G) \lesssim 10~\mathrm{cm^2/V \cdot s}$, where the gate capacitance $C = 11.5~\mathrm{nF/cm^2}$. A regime of zero field effect mobility is also observed, see inset of Fig. 2{\bf B}, where the Ioffe-Regel parameter is $(k_F \lambda)^{-1} \simeq 700$. In contrast, the Ioffe-Regel parameter in Si \cite{Klitzing80} and GaAs/AlGaAs \cite{Tsui82} where QH and FQH was first observed is $(k_F \lambda)^{-1} \simeq 0.02$. The exponential increase in resistance as $T \rightarrow 0~\mathrm{K}$ appears consistent with a variable-range hopping behaviour $R_0 \exp ( (T_0/T)^{n} )$, however we do not find good agreement to either a Mott $n=1/3$ or Efros-Shklovskii $n=1/2$ exponent over the temperature range $1.7~\mathrm{K} < T < 175~\mathrm{K}$. Rather, a pronounced saturation in resistance is observed at the lowest hole density and lowest temperatures, corresponding to the emergence of a zero field effect mobility regime of transport.

The 2-point resistance was measured at $T=575\pm25~\mathrm{mK}$ versus carrier density and in strong magnetic fields, Fig.\ref{fig:fig3}{\bf A}. At high hole densities, a colossal negative magnetoresistance is observed from $B = 0~\mathrm{T}$ through to $B = 45~\mathrm{T}$. A similar magnetoresistance was observed in weakly fluorinated graphene \cite{hong11}. Conversely, in the regime of zero field effect mobility, the resistance is very weakly dependent on magnetic field below a critical field that is itself dependent upon gate voltage. Remarkably, at a gate voltage $V_G = 21~\mathrm{V}$, a sharp transition is observed from an insulating, zero field effect mobility state, to a resistance saturating at a value $R_{2pt} = 12,962~\mathrm{\Omega}$, which is within $0.5\%$ of $h/2e^2$. In a quantum Hall state, the longitudinal four point resistance $R_{xx}\rightarrow 0$ so that the two point resistance approaches the transverse resistance $R_{2pt} \rightarrow |R_{xy}|$. The behaviour of $R_{2pt}$ is fully consistent with a Hall component admitting the quantized value $R_{xy}=h/2e^2$. From the known anomalous QH series of graphene $R_{xy}=\pm h/4e^2(N+1/2)^{-1}$ \cite{Novoselov05,Zhang05}, the hole doping of the sample, and an assumption of preserved spin and valley degeneracies, this QH state corresponds to the $N=-1$ Landau level index and a $\nu=-2$ filling factor. Notably, no Shubnikov-de Haas oscillations in resistance are observed as this QH state is reached at high magnetic field.

\begin{figure}
\includegraphics[scale=.4]{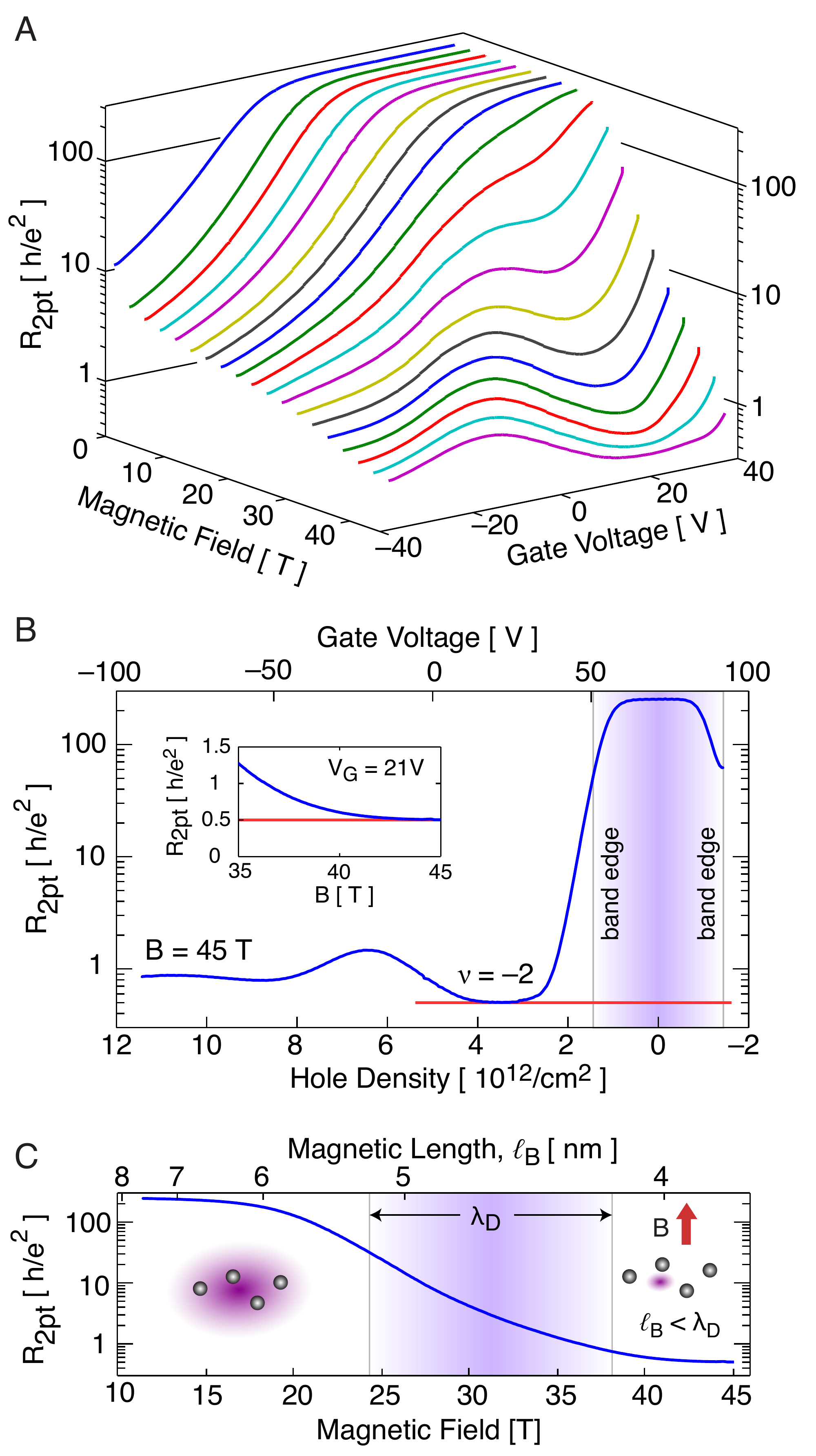}
\caption{Quantum Hall effect in hydrogenated graphene. {\bf A)} The two-point resistance of the hydrogenated graphene sheet is shown versus the magnetic field and the gate voltage. The resistance axis is shown on a logarithmic scale, in units of the resistance quantum $R_{q}=h/e^2$. All data were taken at a temperature of $575\pm25~\mathrm{mK}$. {\bf B)} At 45T, the resistance versus gate voltage (upper {\it x}-axis) and hole density (lower {\it x}-axis), with the red line indicating a Hall plateau at $R=h/2e^2$. The inset shows the resistance at 21V gate voltage approach the quantum Hall state plateau with increasing magnetic field. {\bf C)} Resistance of the hydrogenated graphene at 21V gate voltage plotted versus both the magnetic field (lower {\it x}-axis) and magnetic length $\ell_{B}$ (upper {\it x}-axis). The shaded region indicates the estimated point defect spacing extracted from the Raman spectra. } 
\label{fig:fig3}
\end{figure}

The resistance versus gate voltage at constant $B = 45~\mathrm{T}$, Fig.\ref{fig:fig3}{\bf B}, reveals the charge neutrality point at $V_G = 71\pm 2~\mathrm{V}$ between electron and hole conduction. The hole density at $\nu = -2$ required by the magnetic flux density is $2 e B / h = 2.17 \times 10^{12}\mathrm{/cm^2}$, while the plateau is observed with $3.6 \times 10^{12}\mathrm{/cm^2}$ holes induced from neutrality. The mobile hole valence band edge is thus located at gate voltage $V_G = 51~\mathrm{V}$, providing a signature in electron transport of the formation of a disorder induced gap. Assuming electron-hole symmetry about charge neutrality, we estimate a localized mid-gap state density of $2.9\times10^{12}\mathrm{/cm^2}$.  Gap opening and the presence of mid-gap states have been observed in hydrogenated graphene by ARPES \cite{Grueneis10} and STM \cite{Grueneis12}, with a gap of $\sim 400-600~\mathrm{meV}$ observed for a similar $~0.1-0.3\%$ hydrogen to carbon ratio. Neither the $\nu = -4$ or $\nu = -6$ plateaus are clearly observed, however the weak resistance minimum at $V_G = -51\pm2~\mathrm{V}$ hints to an emerging $\nu=-6$ QH state. Disorder induced broadening of Landau levels may be responsible for the absence of other plateaus and the absence of Shubnikov-de Haas oscillations.

From the perspective of two-dimensional transport in a disordered medium, the emergence of a $\nu=-2$ quantum Hall state at $45~\mathrm{T}$ from a zero-field insulating state induced by neutral point defects can be understood by comparison of length scales, Fig.\ref{fig:fig3}{\bf C}. The magnetic length $\ell_B = \sqrt{\hbar/eB}$ quantifies the magnetic confinement of charge carriers in 2D, independent of material parameters. The insulator-QH transition is here observed with magnetic length spanning $\ell_B=4-6~\mathrm{nm}$, consistent with a crossover from weak to tight magnetic confinement with respect to the point defect spacing $\lambda_D=4.6\pm0.5~\mathrm{nm}$ inferred from Raman spectroscopy. The perspective of an insulator-QH transition arising from magnetic confinement of charge carriers below the mean defect spacing is complementary to the picture of extended states floating in energy that has been previously used to explain insulator-QH transitions \cite{Sankar}.

\begin{figure}
\includegraphics[scale=0.33]{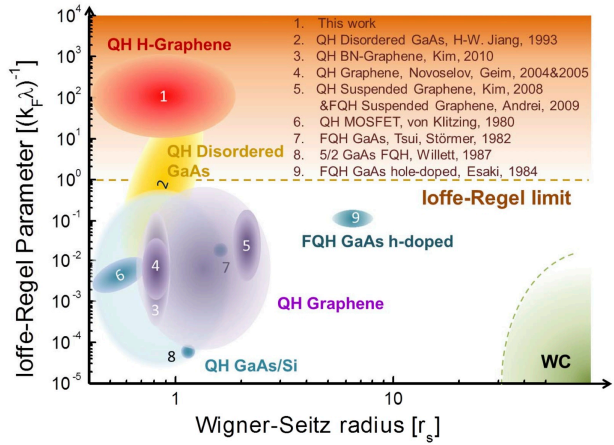}
\caption{ `Ioffe-Regel' - `Wigner-Seitz' diagram for the main families of materials exhibiting the quantum Hall effect. The figure shows the zero-field value of the Ioffe-Regel disorder parameter $(k_{F}\lambda)^{-1}$ plotted versus the Wigner-Seitz radius $r_{s}$, both dimensionless, for the main 2DEG families in which the integer and fractional quantum Hall effect was observed. A brief description of the material system, indexed numerically, is  given in the upper right. The size of the labeled regions is representative of the parameter range from observed mobilities and densities. The dashed line shows the Ioffe-Regel limit where $k_{F}\lambda=1$. The Wigner crystal (WC), thought to occur at large $r_{s}$ and low disorder, is highlighted in green.   }
\label{fig:fig4}
\end{figure}

The emergence of a quantum Hall state from a zero field effect mobility insulating state raises the question of how much disorder can be introduced into a 2D electron gas (2DEG) and still form a quantum Hall state. A diagram summarizing 2DEG systems where the quantum Hall effect (QHE) has been observed is plotted in Fig.\ref{fig:fig4} versus the dimensionless Ioffe-Regel disorder parameter $(k_F \lambda)^{-1}$, with $\lambda$ the transport mean free path, and the dimensionless Wigner-Seitz radius $r_S$ quantifying interaction strength. The preponderance of 2DEGs where the QHE has been observed are low-disorder samples satisfying the Ioffe-Regel criterion (in zero magnetic field) $(k_F \lambda)^{-1}\ll 1$. However, a previous study in disordered  GaAs/AlGaAs heterostructures biased to subthreshold conditions exhibited strong disorder $(k_F \lambda)^{-1} >1$ at low carrier density and a transition from localization to a QH state\cite{Jiang93}. Our observations with hydrogenated graphene pushes the limit of disorder to $(k_F \lambda)^{-1} \gtrsim 500$ where the QHE can still be attained in a strong magnetic field, suggesting that the QHE might be robust to arbitrarily large disorder.

The insulator-QH transition in hydrogenated graphene opens a new regime in energy scales previously unavailable to experiments. The Landau level energy gap about the $N=0$ Landau level is $\Delta E_{0\rightarrow \pm 1} = \sqrt{2}\hbar v_F / \ell_B$ \cite{Zhang05}, giving  $\Delta E_{0\rightarrow \pm 1} = 240~\mathrm{meV}$ at $45~\mathrm{T}$. Previous reports of ARPES of graphene hydrogenated to a $0.1-0.3~\%$ hydrogen-carbon ratio gives an estimated zero-field gap of $E_G \sim 400-600~\mathrm{meV}$ \cite{Grueneis10}. In comparison, the bandgap of GaAs, the choice material for high mobility 2DEGs, is $\sim 1.5~\mathrm{eV}$ which is $\sim 20\times$ larger than the cyclotron energy $\hbar e B / m* = 78 ~\mathrm{meV}$ separating Landau levels at $45~\mathrm{T}$. We propose that the energy scales experimentally accessible in hydrogenated graphene enable the study of the competition between disorder and magnetic confinement, leading to new understanding of the role of disorder in the QHE.

{\bf Acknowledgments}

\begin{small}
We acknowledge useful discussions with A. Grueneis (IFW Dresden), and the outstanding technical assistance of D. Berry, M. Nannini, R. Talbot, J. Smeros, R. Gagnon (McGill) as well as T.P. Murphy, J. Pucci and G. Jones (NHMFL). This work was funded by the Natural Sciences and Engineering Research Council of Canada (NSERC), the Canadian Institute for Advanced Research (CIFAR), the Fonds de Recherche du Qu\'ebec - Nature et Technologies (FRQNT), and the Canada Research Chair program (CRC). A portion of this work was performed at the National High Magnetic Field Laboratory which is supported by NSF Cooperative Agreement No. DMR-0084173, the State of Florida, and the DOE.
\end{small}


%\nocite{*}

%\bibliography{cool_refs}% Produces the bibliography via BibTeX.

%\begin{thebibliography}{10}

\newpage

\preprint{APS/123-QED}

\title{Supplementary Information}% Force line breaks with \\

\author{J. Guillemette$^{1,2}$,  S.S.  Sabri$^{2}$, B. Wu$^{1}$, K. Bennaceur$^{1}$, P.E. Gaskell$^{2}$, M. Savard$^{1}$, P.L. L\'evesque$^{3}$, F. Mahvash$^{2,4}$, A.  Guermoune$^{2,4}$, M.  Siaj$^{4}$, R. Martel$^{3}$, T. Szkopek$^{2\dag}$, and G.  Gervais$^{1\dag \star}$}

\affiliation{$^{1}$Department of Physics, McGill University, Montr\'{e}al, QC,  H3A 2T8, Canada}%Lines break automatically or can be forced with \\

\affiliation{$^{2}$ Department of Electrical and Computer Engineering, McGill University, Montr\'{e}al, QC, H3A 2A7, Canada}%Lines break automatically or can be forced 
\affiliation{$^{3}$ Department of Chemistry, Universit\'e de Montr\'eal, Montr\'{e}al, QC H3C 3J7, Canada}%Lines break automatically or can be forced 

\affiliation{$^{4}$ Department of Chemistry, Universit\'e du Qu\'ebec \`a Montr\'eal, Montr\'{e}al, QC, H3C 3P8, Canada}%Lines break automatically or can be forced \\

\affiliation{$^\dag$both authors contributed equally to this work}
\affiliation{$^\star$corresponding author: gervais@physics.mcgill.ca}
%\date{\today }

%\email{gervais@physics.mcgill.ca}
\date{\today}% It is always \today, today

\maketitle

%\tableofcontents

%\section{Introduction}


%\nocite{*}

%\bibliography{cool_refs}% Produces the bibliography via BibTeX.

%\begin{thebibliography}{10}

\section{\Large SUPPLEMENTARY INFORMATION}

\vspace*{5mm}

\section{Device fabrication}

Graphene monolayers were grown by chemical vapour deposition (CVD) on electropolished $25~\mu m$ thick copper foils with a methane precursor. Complete CVD growth procedures are described in detail in previous work \cite{guermoune11}, including the dependence of material quality on growth parameters. Graphene monolayers were transferred to oxidized silicon substrates using a thin poly-methylmethacrylate handle layer, sacrificial etching of the Cu foil in a 0.1M (NH$_4$)$_2$S$_2$O$_8$ room temperature solution. The substrates were $500~\mathrm{\mu m}$ thick As doped Si with $1-5~\mathrm{m\Omega-cm}$ room temperature resistivity. A chlorinated, dry thermal oxide of 300~nm thickness was used as the back-gate dielectric. After graphene transfer, a shadow mask was used to deposit Ti/Au (2~nm/50~nm) Ohmic contacts 3~mm apart. A scribe was used to manually isolate individual graphene squares on the substrate surface.

Hydrogenation was performed in a UHV chamber with a base pressure in the low $10^{-9}$~Torr range. A custom-built capillary atomic hydrogen source \cite{Bischler93} was used. A tungsten capillary was heated to $>1800~\mathrm{K}$ by electron bombardment, with $\sim12~\mathrm{W}$ power at $\sim600~\mathrm{V}$ acceleration voltage. A beam of thermally cracked hydrogen was produced upon introduction of molecular hydrogen to the capillary through a leak valve at an approximate chamber pressure of $10^{-5}-10^{-6}$ Torr. The graphene samples were mounted $\sim10~\mathrm{cm}$ away from the hydrogen source, with a shutter used to precisely control device exposure.  All devices were subjected to a 1 hour thermal anneal at 400~K prior to hydrogen exposure to degas the sample surface. Between hydrogen doses, {\it in situ} measurement  of electrical resistance was performed, allowing the effect of hydrogenation on electrical resistance to be monitored.

 \section{Raman spectroscopy}

\begin{figure}
\includegraphics[scale=0.5]{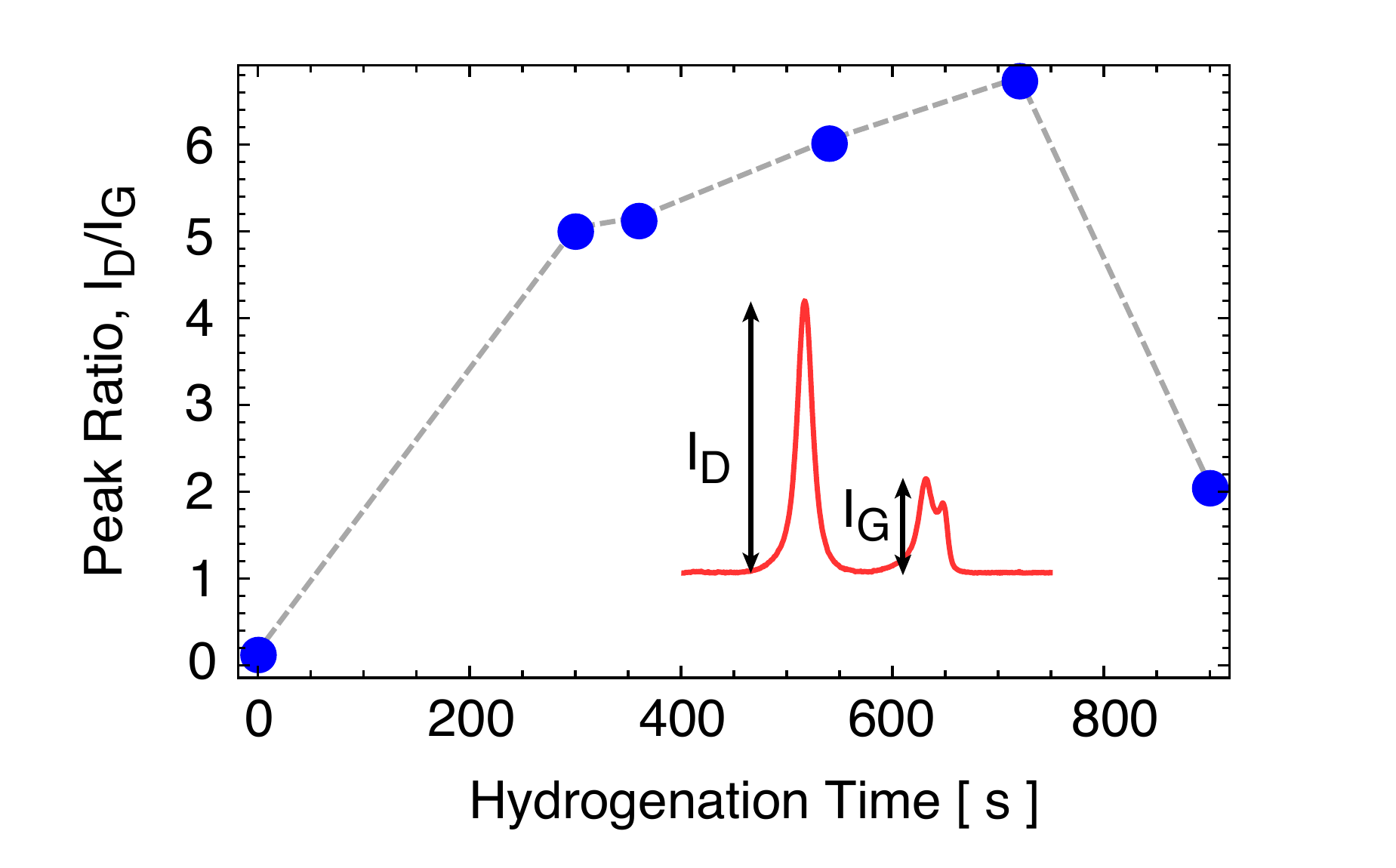}
\caption{ Raman analysis of hydrogenated graphene. The ratio of Raman D peak intensity $I_D$ to G peak intensity $I_G$ versus hydrogenation time, obtained with a pump wavelength $\lambda = 633~\mathrm{nm}$. Saturation in $I_D/I_G$ corresponds to an approximate defect spacing $\lambda_D \sim 4~\mathrm{nm}$. Electron transport measurements were performed on the sample hydrogenated for 360~s. }
\label{fig:figS1}
\end{figure}

Raman spectroscopy was performed on CVD graphene and hydrogenated CVD graphene with a $\lambda = 532~\mathrm{nm}$ pump laser of 20~mW incident power and a $\lambda = 633~\mathrm{nm}$ pump laser of 20~mW incident power, both through a 40$\times$ objective. All measurements were performed with graphene on oxidized silicon, with 4 distinct spatial regions probed. With the $\lambda = 532~\mathrm{nm}$ pump, the Raman peaks observed in CVD graphene (Fig. 1{\bf A}) were D (1347~cm$^{-1}$), G (1588~cm$^{-1}$), G* (2462~cm$^{-1}$), G' (2685~cm$^{-1}$) and G" (3255~cm$^{-1}$). The D peak was more intense relative to G in the hydrogenated graphene sample used for high-field transport measurements, and new Raman peaks associated with increased disorder \cite{Jorio11} emerged: G+D (2942~cm$^{-1}$) and D' (1628~cm$^{-1}$). The evolution of the ratio of D peak intensity ($I_D$) to G peak intensity($I_G$) versus hydrogenation time of six CVD graphene samples is plotted in Fig.\ref{fig:figS1}. A non-monotonic behaviour in $I_D/I_G$ versus hydrogen dose is found, similar to the evolution of $I_D/I_G$ versus Ar$^+$-ion bombarded graphene \cite{Lucchese10}. Under the assumption of similar point-defect induced Raman spectra by hydrogenation as by Ar$^+$-ion bombardment, the observed saturation in $I_D/I_G$ corresponds to a mean defect spacing $\lambda_D \sim 4~\mathrm{nm}$.

The hydrogenated sample (\#HG30) used for high-field (45 T) electron transport measurements reported here was analyzed with greater precision by Raman spectroscopy within 4 days of electron transport measurement, to minimize the potential effect of environmental exposure between electron transport and Raman spectroscopy. To estimate the hydrogen induced defect density, we apply the same phenomenological model for point-defect induced disorder as has been successfully applied to understanding the correlation between defect spacing $\lambda_D$ and Raman peak ratio $I_D/I_G$ in Ar$^+$-ion bombarded graphene \cite{Lucchese10}. The phenomenological mapping between $I_D/I_G$ and $\lambda_D$ is, $\frac{I_D}{I_G} = C_A  \frac{r_A^2 - r_S^2}{r_A^2-2r_S^2} \left[ \exp\left( \frac{-\pi r_S^2}{\lambda_D^2} \right) - \exp\left( \frac{-\pi \left( r_A^2 - r_S^2 \right) }{\lambda_D^2} \right) \right]+ C_S  \left[ 1 - \exp \left( \frac{-\pi r_S^2}{\lambda_D^2} \right) \right] $
where $C_A = 4.2$, $C_S = 0.87$, $r_A = 3.0~\mathrm{nm}$ and $r_S = 1.0~\mathrm{nm}$ for the pump wavelength $\lambda = 514~\mathrm{nm}$. Our Raman spectra give the ratio $I_D/I_G = 2.88\pm0.06$ at $\lambda = 532~\mathrm{nm}$, with the uncertainty determined by the standard deviation between different measurement points on the sample. In the absence of a pump wavelength dependence of $I_D/I_G$ for point defects, we conservatively estimate the error in $\lambda_D$ as follows.
The known $I_D/I_G$ scaling of $\lambda^{4}$ for edge defects \cite{Jorio11} gives an estimated correction $I_D/I_G(514~\mathrm{nm}) = (514/532)^4 = I_D/I_G(532~\mathrm{nm}) = 2.51 \pm 0.05$. We conservatively take the widest error admitted by the pump wavelength scaled and unscaled Raman ratios $I_D/I_G = 2.70 \pm 0.24$, giving an estimated point-defect spacing from Eq. 1 of $\lambda_D = 4.6 \pm 0.5 ~\mathrm{nm}$.
 
 \section{Electron transport measurement}
 
\begin{figure}
\includegraphics[scale=0.33]{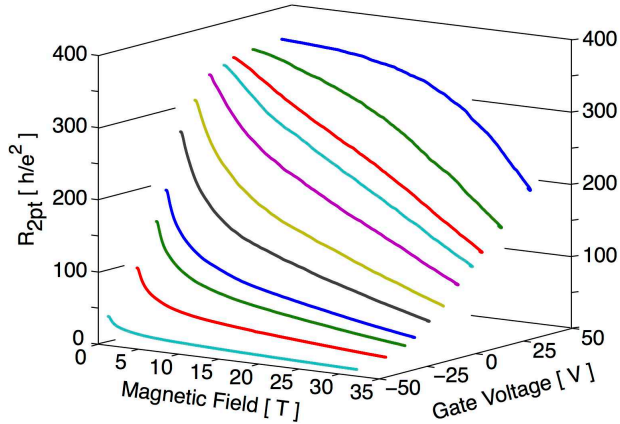}
\caption{Colossal negative magnetoresistance of hydrogenated graphene. Resistance versus magnetic field (up to 33 T) and gate voltage of sample \#HG18 is among five that exhibited colossal negative magnetoresistance. These measurements, performed on a different sample and during a different magnet run, are fully consistent with the measurements presented for sample \#HG30 described in the main body of the manuscript.   }
\label{fig:figS2}
\end{figure}

The hydrogenated graphene devices were mounted on ceramic carriers, in turn mounted on G10 sample holders, that were then mounted to either a pumped $^4$He or $^3$He refrigerator. The samples were cooled in either the gas or liquid phase of helium. Preliminary characterization was performed with a 9~T solenoid (McGill). All electrical measurements on hydrogenated graphene were performed using standard lock-in techniques at 13~Hz and at a low current excitation $I_{exc} \lesssim 10$~nA. Gate voltage was applied with a DC voltage source, and gate leakage current was monitored for all measurements with a DC ammeter. The gate leakage resistance was at minimum $2~\mathrm{G\Omega}$. High-field measurements were performed on samples using the both a 33~T resistive magnet and the hybrid superconducting-resistive magnet of the NHMFL facility in Tallahassee. With the hybrid system, a DC magnetic field of 45~T was applied normal to the graphene surface with a superconducting outsert coil generating up to 11.5~T, and an additional 34~T field generated by a resistive insert magnet. Colossal negative magnetoresistance was observed in five distinct samples, with the resistance versus magnetic field and gate voltage of sample \#HG18 given in Fig.\ref{fig:figS2}.

\end{document}